\documentclass[12pt,pre,aps,reprint,showpacs]{revtex4-1}

\usepackage{graphicx}
\usepackage{amsfonts}
\usepackage{amsmath}
\usepackage{mathtools}
\usepackage{float}
\usepackage{epsfig}
\usepackage{color}
\usepackage{gensymb}
\usepackage{multirow}
\usepackage[mathscr]{euscript}
\usepackage{pifont}
\usepackage{longtable}

\newcommand{\e}[1]{\ensuremath{\times 10^{#1}}}
\newcommand{\ee}[1]{\ensuremath{10^{#1}}}

\begin{document}

\title{Classical many-particle systems with unique disordered ground states}

\author{G. Zhang}


\affiliation{\emph{Department of Chemistry}, \emph{Princeton University},
Princeton NJ 08544 }

\author{F. H. Stillinger}


\affiliation{\emph{Department of Chemistry}, \emph{Princeton University},
Princeton NJ 08544}

\author{S. Torquato}

\email{torquato@electron.princeton.edu}

\affiliation{\emph{Department of Chemistry, Department of Physics,
Princeton Institute for the Science and Technology of
Materials, and Program in Applied and Computational Mathematics}, \emph{Princeton University},
Princeton NJ 08544}

\begin{abstract}

Classical ground states (global energy-minimizing configurations) of many-particle systems are typically unique crystalline structures, implying zero enumeration entropy of distinct patterns (aside from trivial symmetry operations).
By contrast, the few previously known disordered classical ground states of many-particle systems are all high-entropy (highly degenerate) states. 
Here we show computationally that our recently-proposed ``perfect-glass'' many-particle model 
[Sci. Rep., {\bf 6}, 36963 (2016)] possesses disordered classical
ground states with a zero entropy: a highly counterintuitive situation. 
For all of the system sizes, parameters, and space dimensions that we have numerically investigated, the disordered ground states are unique such that they can always be superposed onto each other or their mirror image.
At low energies, the density of states obtained from simulations matches those calculated from the harmonic approximation near a single ground state, further confirming ground-state uniqueness.
Our discovery provides singular examples in which entropy and disorder are at odds with one another.
The zero-entropy ground states provide a unique perspective on the celebrated Kauzmann-entropy crisis in which the extrapolated entropy of a supercooled liquid drops below that of the crystal.
We expect that our disordered unique patterns to be of value in fields
beyond glass physics, including
applications in cryptography as pseudo-random functions with tunable computational complexity.
\end{abstract}

\maketitle

\section{Introduction}
The classical ground states of many-particle systems are typically crystals consisting of periodically replicated energy-minimizing local geometries with high symmetry.
The ability for the particles to attain and display long-range order (Bragg diffraction) becomes the likely procedure for those models to attain their ground state.
A specific system at a fixed density usually possesses a unique crystal ground state, aside from trivial symmetry operations.
Therefore, the ``enumeration entropy''
\begin{equation}
\mathscr S_E=k_B\ln\Omega_E，
\end{equation}
is zero for such ground states. Here $\Omega_E$ is the number of {\it distinct} accessible structures and $k_B$ is the Boltzmann constant.
 
\begin{figure}
\includegraphics[width=0.4\textwidth]{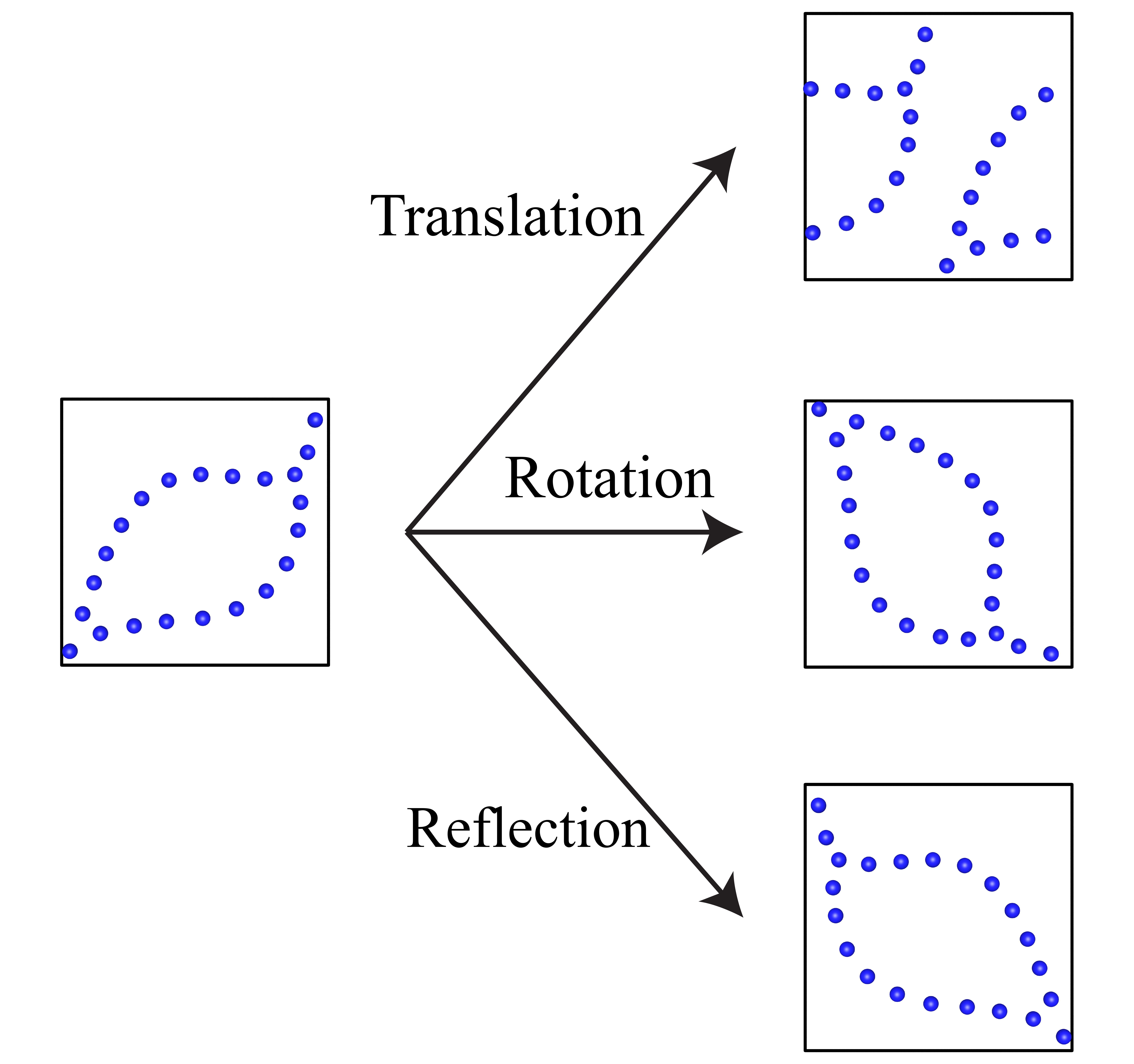}
\caption{Illustration of the three pattern-preserving symmetry operations. Two configurations have the same pattern if they are related to each other through any combination of these three symmetry operations.}
\label{fig:SymmetryCartoon}
\end{figure}

The fact that ground states of many-body systems can be disordered have intrigued condensed-matter physicists.
Although quantum effects are the cause of ground-state disorder in many systems (for example, helium under normal pressure \cite{feynman1955chapter} and certain spin systems \cite{kennedy1988two, sachdev1992kagome, nakamura2002order, chen2016fermionic}), classical systems can also have disordered ground states \cite{edwards1975theory, hansen1990theory, marinari1994replica, marinari1994replica2, uche2006collective, batten2008classical, batten2008classical,  zachary2011anomalous, torquato2015ensemble}. A ground state of a classical many-particle or spin system is simply a global minimum of the potential energy. For classical many-particle systems in Euclidean spaces, all known examples of disordered ground states possess high enumeration entropy, in the sense that there exists an uncountable collection of geometrically inequivalent ground-state configurations. Here, ``inequivalent'' configurations are those that are not related to each other by trivial symmetry operations, which includes translations, rotations, and reflections (illustrated in Fig.~\ref{fig:SymmetryCartoon}).
Such examples include equilibrium hard-sphere systems away from jammed states \cite{hansen1990theory} and particles interacting with ``stealthy'' and related collective-coordinate potentials \cite{uche2006collective, batten2008classical, batten2008classical, zachary2011anomalous, torquato2015ensemble, zhang2015ground}.
While the former situation is trivial in that any nonoverlapping configuration counts as a ground state, the latter systems are less so because certain nonlinear constraints are imposed on the configuration.
Depending on the specific constraints, the latter interactions can create stealthy systems \cite{torquato2015ensemble}, ``super-ideal gases'' \cite{batten2008classical}, ``equi-luminous materials'' \cite{batten2008classical}, as well as other unusual ground states \cite{uche2006collective, zachary2011anomalous}. 

It it natural to expect that the entropy of these disordered ground states is large and extensive for two reasons. First, entropy has often been associated with the amount of disorder in a system.
It was not until 1949 that Onsager realized that entropy and disorder
are not always directly related to one another by showing that the entropy of a fluid of hard needles can increase when the needles tend to align with 
one another, thereby increasing the orientational order of the system \cite{onsager1949effects}. 
Hard spheres also undergo an entropically driven disorder-order phase transition at sufficiently
high densities \cite{alder1957phase, frenkel1999entropy}. 
Second, as the aforementioned examples illustrate, the tendency for ground states to be disordered is caused by the nature of the interactions, which allows certain individual or collective displacements of particles without causing any change in the energy. A ground-state configuration can thus move in these unconstrained directions of the configuration space, and thus become pattern-degenerate with large and extensive entropy. Here we define a set of ground states to be pattern-unique if all of the ground state structures are equivalent, and pattern-degenerate otherwise \footnote{It should be stressed that this paper focuses on classical many-particle systems.  If one includes spin systems (or equivalently, lattice-gas systems), which by definition lack continuous deformations, there are known examples of unique or pattern-unique classical ground states.
In the ``low-correlation'' spin model \cite{marinari1994replica} and a subsequent simplified model \cite{marinari1994replica2}, the ground state is pattern-unique for a small fraction of system sizes but degenerate for other system sizes, while the degeneracy in the infinite-system-size limit is uncertain. (Although neither Ref.~\onlinecite{marinari1994replica} nor Ref.~\onlinecite{marinari1994replica2} explicitly commented on the ground state degeneracy, we subsequently enumerated all possible spin configurations for 10-20 sites, and found that the ground states of the one-dimensional model described in Ref.~\onlinecite{marinari1994replica} are degenerate, even after removing trivial translations and reflections, except for number of sites $N_s=11$ and $N_s=15$. Ref.~\onlinecite{marinari1994replica2} stated that its model has the same ground states as the model described in Ref.~\onlinecite{marinari1994replica}.) Another example, the well-known spin-glass models, has an unambiguously pattern-unique disordered ground state \cite{edwards1975theory}, but does so in a trivial way: the interaction is disordered (different for each pair of neighboring sites), causing the disordered ground state.}.

In this paper, we demonstrate that our recently-proposed ``perfect-glass'' many-particle model \cite{zhang2016perfect} surprisingly possesses classical ground states that are counterintuitively disordered with zero enumeration entropy.
Perfect glasses are distinguished from normal glasses and other amorphous solids in that they are by construction hyperuniform (anomalously suppress large-scale density fluctuations), as defined by a static structure factor that tends to zero in the infinite-wavelength limit \cite{torquato2003local}; 
see Refs.~\onlinecite{ torquato2016hyperuniformity,Ma17,Ch17,He17} for recent developments on disordered hyperuniform systems.
Moreover, since perfect glasses can never crystallize or quasicrystallize at zero or any positive temperature \cite{zhang2016perfect}, they circumvent the Kauzmann entropy crisis in which the extrapolated entropy of a supercooled liquid drops below that of the crystal \cite{kauzmann1948nature}.
By contrast, traditional glasses have been venerably understood as liquids
kinetically arrested from cooling that are metastable with respect to a crystal \cite{angell1988perspective, sastry2001relationship, shintani2006frustration, ediger2012perspective, gupta2015validity}.
The  unique disordered ground states of perfect-glass models are to be contrasted with zero-entropy crystals and quasicrystals that possess high symmetry and long-range translational and/or rotational order.
Thus, these disordered ground states can be a fertile area for future research in disciplines beyond physics. 

It is noteworthy that unlike spin-glass models \cite{edwards1975theory}, perfect-glass interactions treat all particles equally and thus do not introduce disorder by the intrinsic random nature of the  interactions; unlike the low-correlation spin model \cite{marinari1994replica, marinari1994replica2}, the ground state is pattern-unique for all finite system sizes we have studied, and is therefore expected to be pattern-unique in the infinite-system-size limit.

The rest of the paper is organized as follows: In Sec. II, we provide basic
definitions. In Sec. III, we numerically show that perfect-glass
ground states are pattern unique by enumerating the minima of the potential energy surface.
In Sec. IV, we compute the density
of states of perfect glasses as a function of the 
potential energy with two different approaches: one assuming ground-state pattern uniqueness, and another without such an assumption.
We show that the results from these two different approaches are in excellent agreement, which confirms the ground-state pattern uniqueness.
In Sec. V, we provide conclusions
and discuss the broader implications of our findings.




\section{Basic definitions}
For a single-component system with $N$ particles located at positions $\mathbf r_1, \mathbf r_2, \cdots, \mathbf r_N$, in a simulation box of volume $V$ subjected to periodic boundary conditions in $d$-dimensional Euclidean space $\mathbb R^d$, the static structure factor is defined as
\begin{equation}
S(\mathbf k)=\frac{\Big|\sum_{j=1}^N \exp(-i \mathbf k \cdot \mathbf r_j)\Big|^2}{N},
\end{equation}
where $i$ is the imaginary unit and $\mathbf k$ is a $d$-dimensional wavevector (which must be a linear combination of integer multiples of the reciprocal lattice vectors of the simulation box).

\begin{figure}
\includegraphics[width=0.5\textwidth]{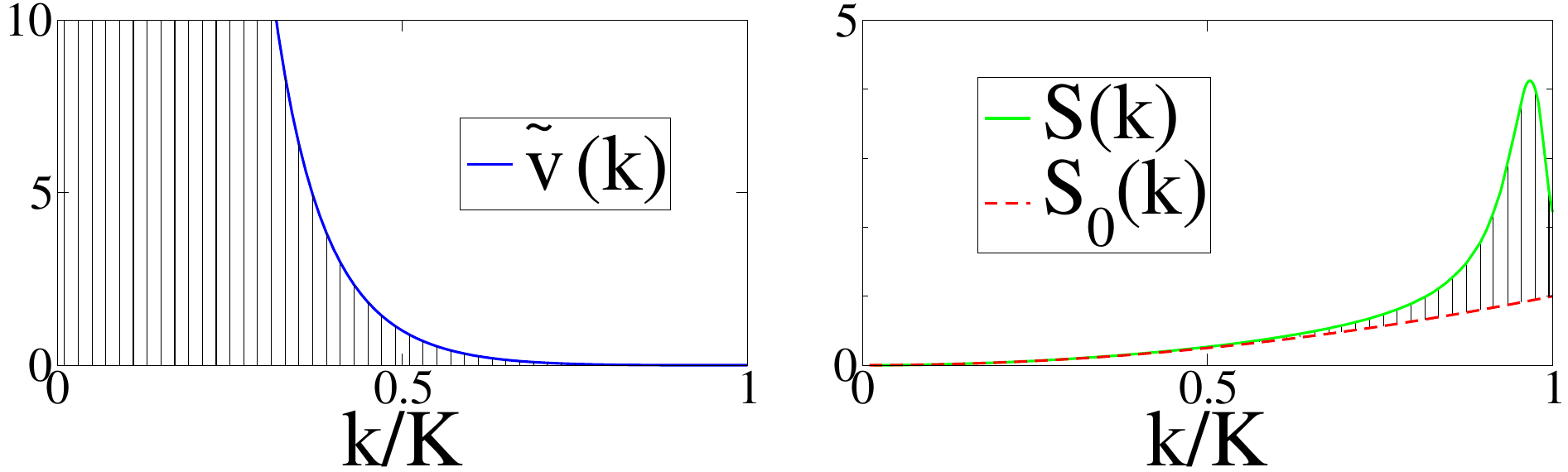}
\caption{Shaded-area illustration of the two multiplicative contributions of the potential energy, defined in Eq.~(\ref{eq:PGpot}).
Notice that even though the target structure factor $S_0(k)$ is monotonic in $k$, the actual structure factor $S(k)$ deviates from it and develops peaks, the first of which is shown here.
}
\label{fig:TargetCartoon}
\end{figure}

The perfect-glass interaction potential \cite{zhang2016perfect} has either
a direct-space or Fourier-space representation. In the latter case, we have
\begin{equation}
\Phi(\mathbf r_1, \mathbf r_2, \cdots, \mathbf r_N) = \sum_{0<|\mathbf k|<K} {\tilde v}(\mathbf k)[S(\mathbf k)-S_0(\mathbf k)]^2.
\label{eq:PGpot}
\end{equation}
This interaction is designed to constrain the static structure factor, $S(\mathbf k)$ to be equal to a target function $S_0(\mathbf k)$, for all wave vectors $\mathbf k$ within a certain distance $K$ from the origin; and assigns energy penalties, adjusted by a weight function
${\tilde v}(\mathbf k)$, if such constraints are violated. 
Following Ref.~\onlinecite{zhang2016perfect}, we use $S_0(\mathbf k)=|\mathbf k|^\alpha$ and ${\tilde v}(\mathbf k)=(K/|\mathbf k|-1)^3$, where $\alpha$ is a positive parameter we can choose freely.
The two multiplicative factors in the summand of Eq.~(\ref{eq:PGpot}) are illustrated in Fig.~\ref{fig:TargetCartoon}.
In general, other forms of $S_0(\mathbf k)$ and ${\tilde v}(k)$ may also be used, but these particular forms were chosen to realize hyperuniformity.
The direct-space representation
of the perfect-glass potential (\ref{eq:PGpot})
involves a sum of two-body, three-body, and four-body interactions \cite{uche2006collective}.

We define $\chi$ to be the ratio of the number of constrained degrees of freedom to the number of independent degrees of freedom, $d(N-1)$ \cite{zhang2016perfect}. When $\chi$ is larger than unity, the system runs out of degrees of freedom and becomes glassy, {\it i.e.,} develops a complex energy landscape with multiple energy minima, and a positive shear modulus \cite{zhang2016perfect}. This model completely banishes crystalline structures at any nonnegative temperature, since the existence of Bragg peaks would make the potential energy infinite \cite{zhang2016perfect}.

\section{Enumeration of the energy minima}
We numerically study the classical ground states of the perfect-glass interactions and demonstrate their pattern uniqueness by showing that the enumeration entropy, defined by relation (1), is zero. 
We minimize the potential energy,
using the low-storage Broyden–Fletcher–Goldfarb–Shanno algorithm \cite{nocedal1980updating, liu1989limited, nlopt},
starting from random initial configurations, to find local minima of the potential energy surface. 
A random local minimum of the potential energy surface is deemed to be reached once the energy minimization routine finishes with a stringent tolerance of $\delta \Phi=\ee{-11}$. Therefore, by repeating this process a sufficient number of times, we expect to find the global minimum of the potential energy surface. 
After $\ee{7}$ to $\ee{9}$ independent energy minimization trials, a lowest energy is achieved at least $10$ times, but often more than $\ee{3}$ times (see Appendix A for details). 
Presumably, this is the ground state energy. Subsequently, we compare the ground-state configurations for pattern uniqueness.
A particular ground-state configuration is taken to be a comparator, and then we compare it to every other ground-state configuration.
Using an algorithm detailed in Appendix B, we attempt to find a translation, a rotation, and/or a reflection so that after these symmetry operations the comparator superposes onto the other ground state. After these symmetry operations are performed, if each particle in the comparator is within $\ee{-5}L$ distance to a particle in the other ground state, then the two ground states are deemed to have the same pattern. Here $L$ denotes the side length of the simulation box. The ground state is considered pattern-unique if all of the ground-state configurations have the same pattern as the comparator.

\begin{figure}
\raisebox{-0.5\height}{\includegraphics[width=0.5\textwidth]{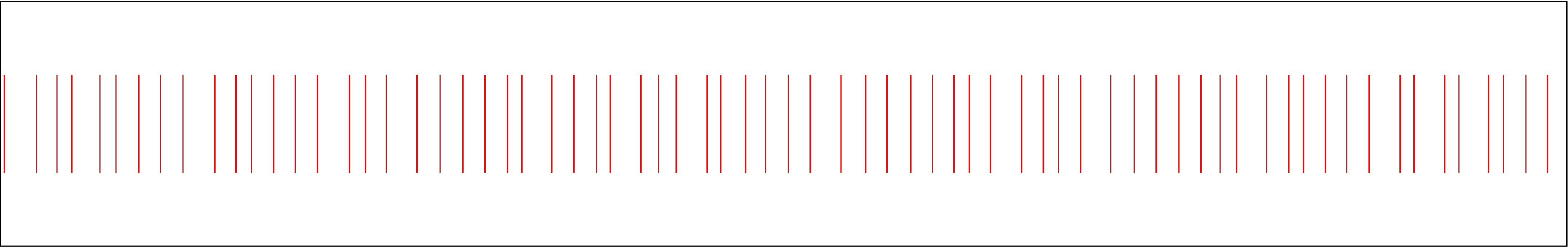}}
\raisebox{-0.5\height}{\includegraphics[width=0.23\textwidth]{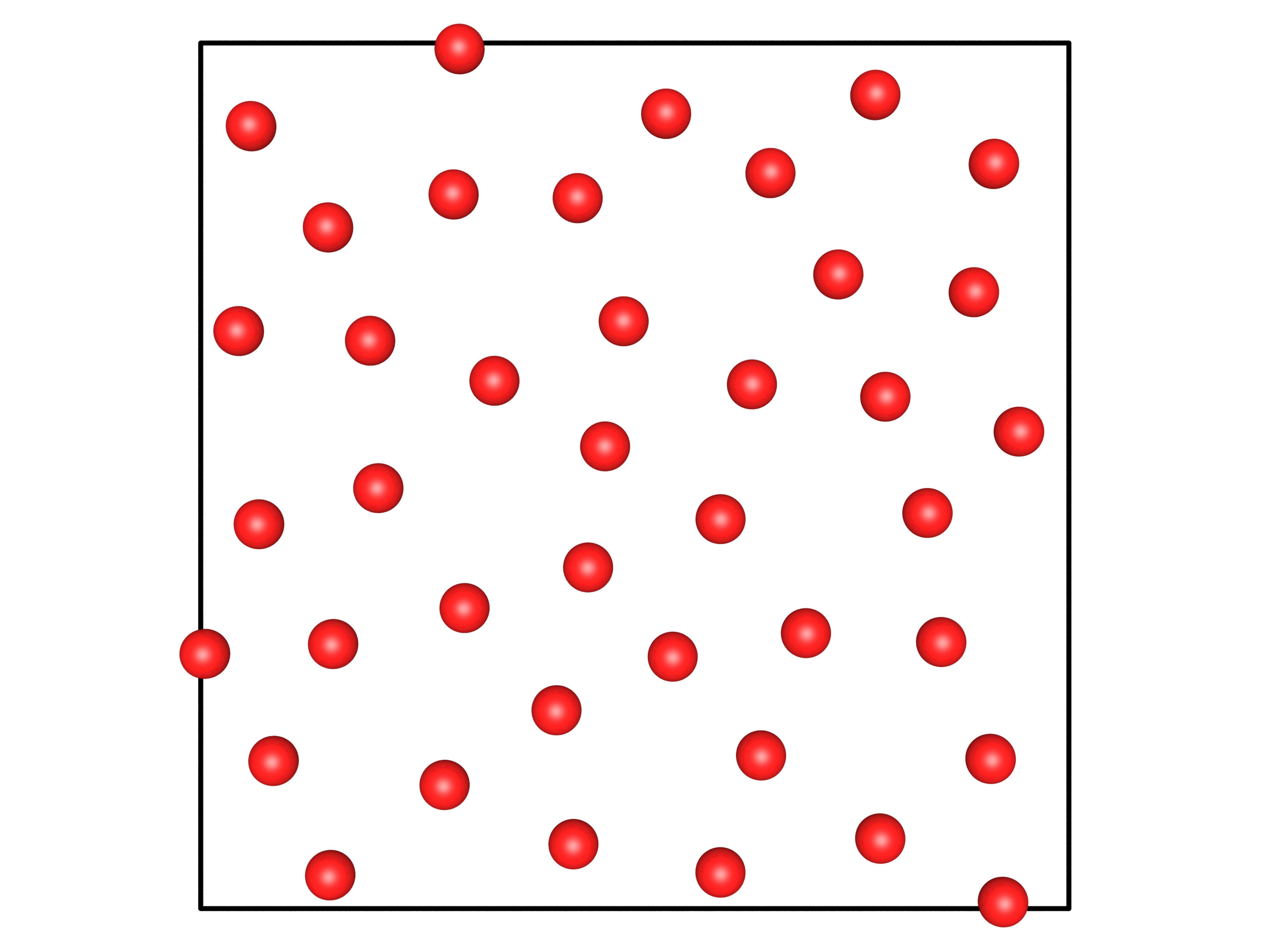}}
\raisebox{-0.5\height}{\includegraphics[width=0.23\textwidth]{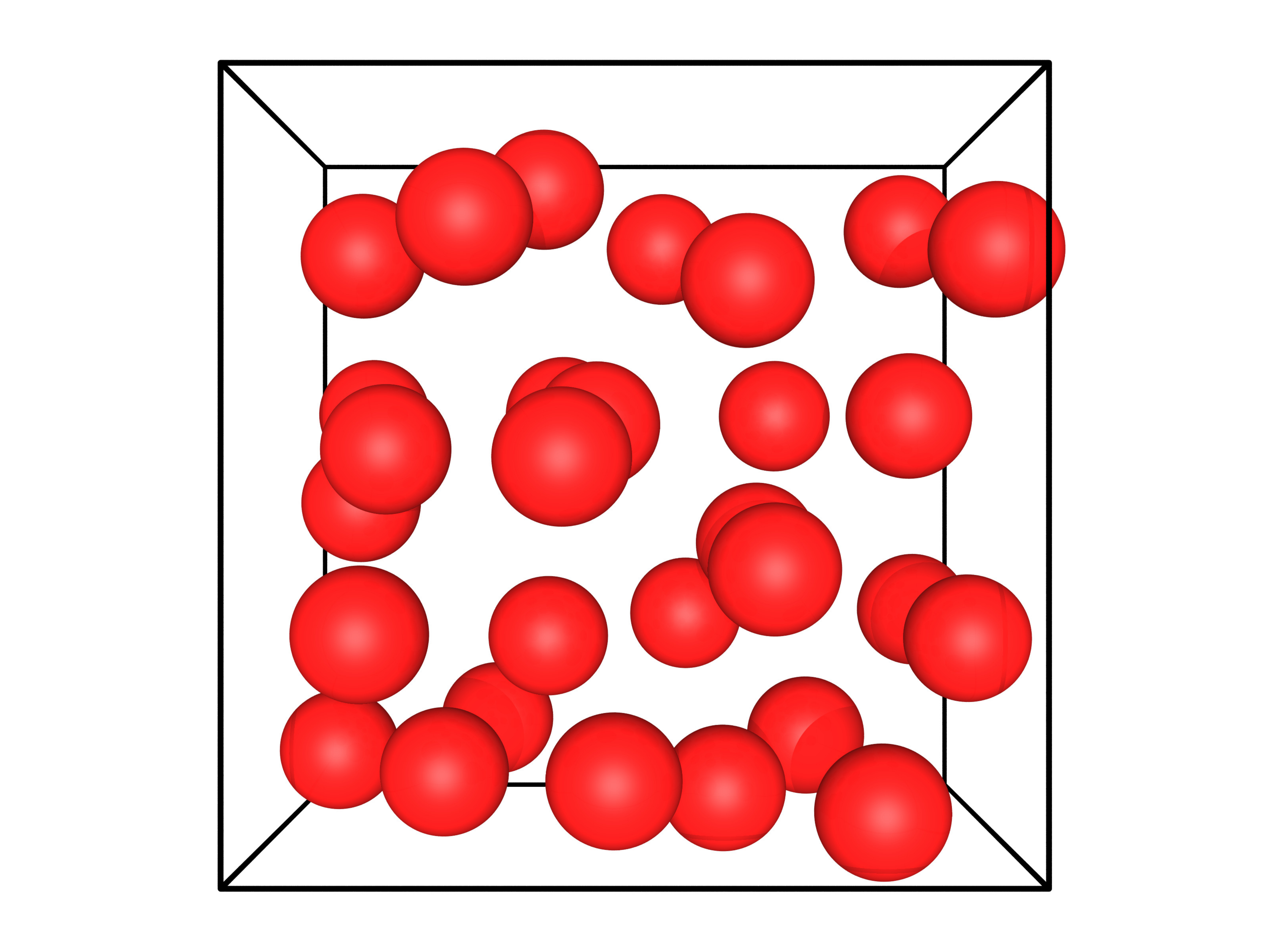}}
\caption{The disordered unique ground states of the perfect-glass potential for (top) $d=1$, $\alpha=6$, $\chi=1.75$, and $N=70$; (bottom left) $d=2$, $\alpha=6$, $\chi=1.87$, and $N=40$; and (bottom right) $d=3$, $\alpha=6$, $\chi=1.75$, and $N=30$. 
These figures illustrate a point presented in Ref.~\onlinecite{zhang2016perfect}, namely, the particles experience a pair repulsion that is clearly observed when one calculates the pair correlation function.}
\label{fig:PGGround}
\end{figure}

We studied a total of 60 different combinations of parameters ($d$, $\alpha$, $\chi$, and $N$); see Appendix A for a complete list. These cases cover wide ranges of $N$ (between 10 and 70, including both prime $N$'s and composite $N$'s), $\alpha$ (between 0.5 and 6), and $\chi$ (between 1.7 and 2), in one, two, and three dimensions.  For all cases, the ground state was found to be disordered and pattern-unique. The discovered ground states of the largest $N$ cases in the first three space dimensions are presented in Fig.~\ref{fig:PGGround}.

\begin{figure*}
\raisebox{-0.5\height}{\includegraphics[width=0.32\textwidth]{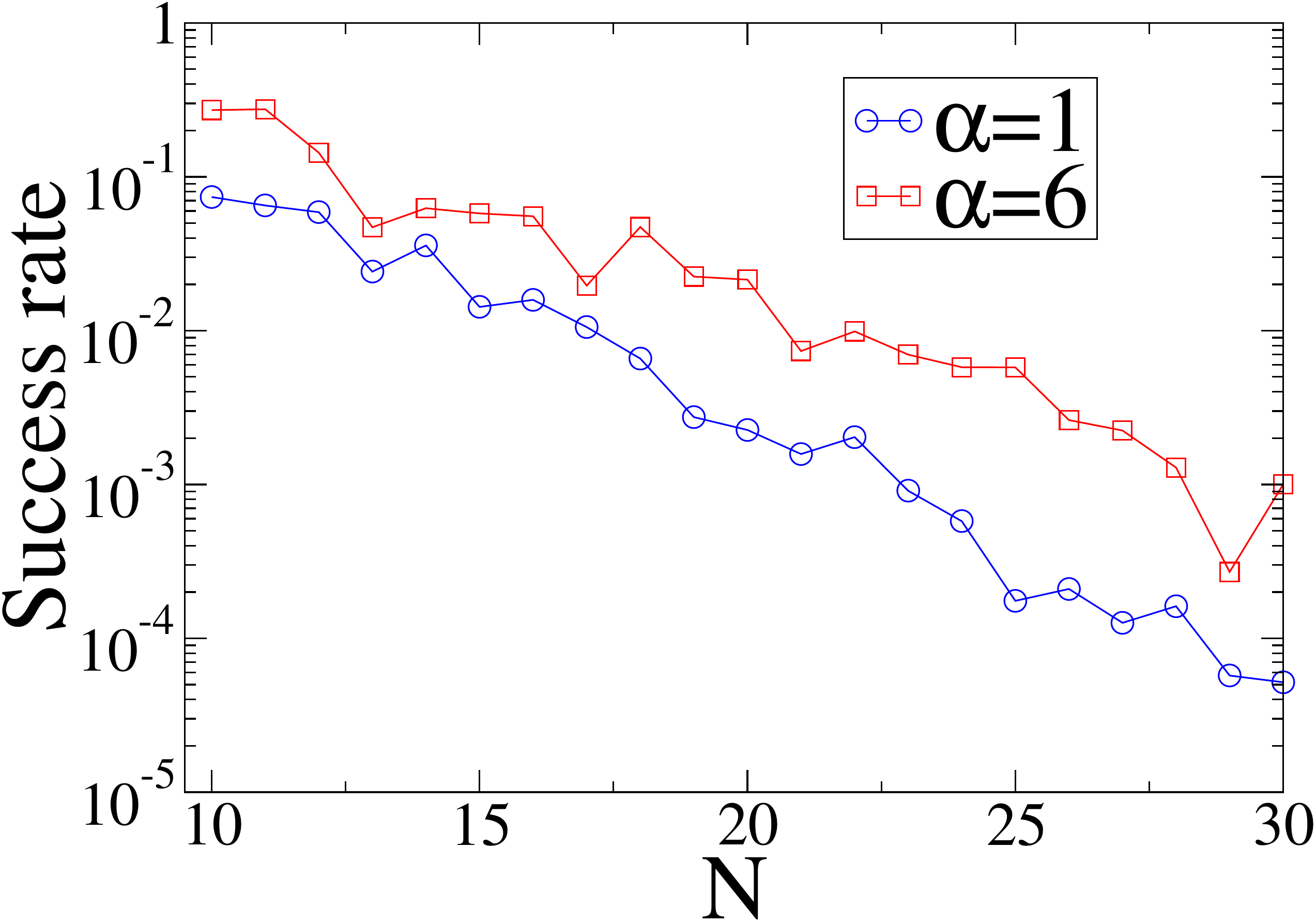}}
\raisebox{-0.5\height}{\includegraphics[width=0.32\textwidth]{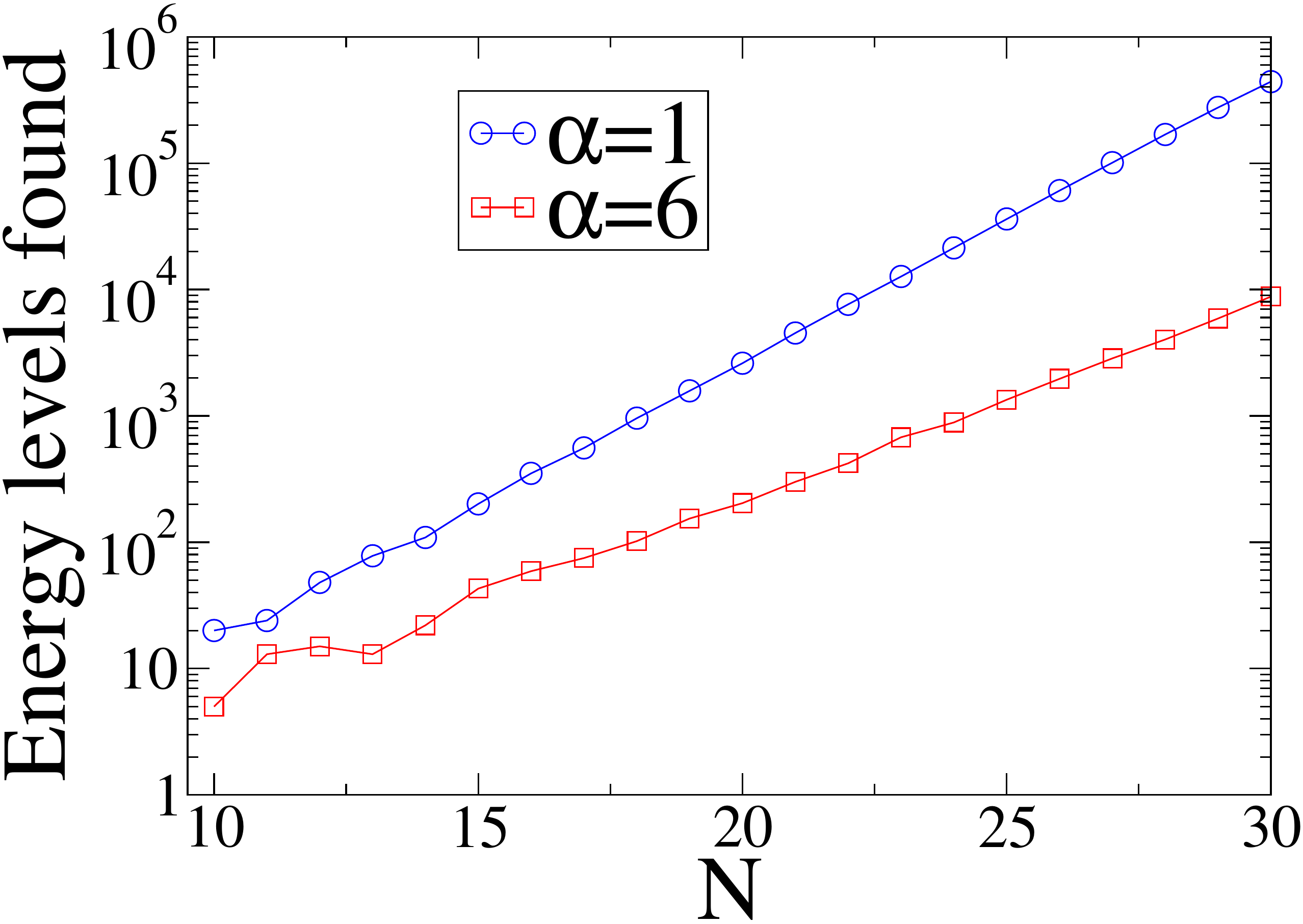}}
\raisebox{-0.5\height}{\includegraphics[width=0.32\textwidth]{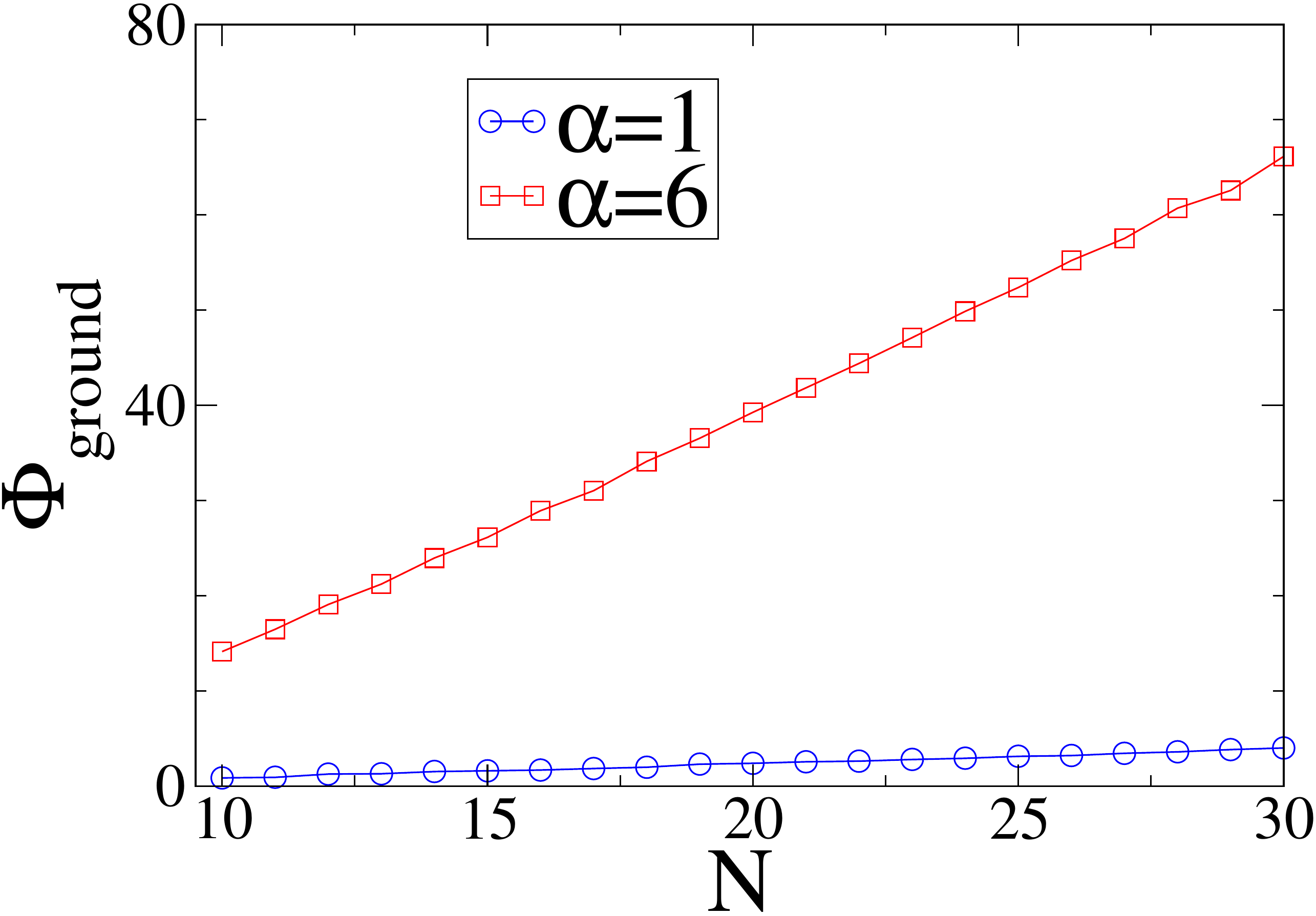}}
\caption{(left) The probability of finding the ground states by energy minimization for $d=1$, $\chi=2.00$, $\alpha=1$ and 6, and $10\le N\le 30$. (middle) The number of distinct energy local minima found by $\ee{7}$ repeated energy minimizations for the same systems. (right) The ground-state energy of the same systems.}
\label{fig:PGStatistics}
\end{figure*}

Besides the ground states, we also study other minima of the potential energy surface. As Fig.~\ref{fig:PGStatistics} shows, as $N$ increases, the success rate (the probability that one finds the ground state through an energy minimization trial) decreases exponentially, and the number of discovered energy minima increases exponentially. This exponential rise of the number of higher minima is in agreement with what one has topographically for real glass formers \cite{stillinger2015energy}. Compared to the $\alpha=1$ case, the $\alpha=6$ case possesses a higher success rate and fewer distinct energy levels. This is also expected because as we have discovered earlier, increasing $\alpha$ increases geometrical order in these glasses \cite{zhang2016perfect}. Finally, Fig.~\ref{fig:PGStatistics} also shows that the ground state energy is roughly proportional to $N$ for both $\alpha$ values we presented.


\section{Calculation of the density of states}
To further confirm ground-state uniqueness, we have also performed Wang-Landau Monte Carlo (WLMC) simulations on a perfect-glass system with $d=2$, $\alpha=1$, $\chi=1.89$, and $N=10$. The WLMC algorithm allows one to calculate the density of states $g(E)$ as a function of the potential energy \cite{wang2001efficient} (or equivalently, the hyper-area of an iso-energy surface in the configuration space). Alternatively, for energy values very close to the ground state, one could also calculate $g(E)$ from the eigenvalues of the Hessian matrix by treating the system as an harmonic oscillator around the ground state.
As detailed below, after considering the aforementioned trivial symmetry operations, we find very good agreement between the calculated $g(E)$'s from these two approaches, further verifying ground-state uniqueness.

\subsection{Density of states, $g(E)$, from the harmonic approximation}
In this subsection, we calculate $g(E)$ using the harmonic approximation.
For a $d$-dimensional configuration of $N$ particles, the configuration space is $dN$-dimensional. Of these $dN$ directions of the configuration space, $d$ directions correspond to translations of the whole configuration, which cause no energy change. The other $d(N-1)$ directions correspond to deformations, which generally change the potential energy. Near the classical ground state, such changes can be quantified by the eigenvalues of the Hessian matrix, $\lambda_1$, $\lambda_2$, $\cdots$, $\lambda_{d(N-1)}$. Let $E$ denote an energy that is slightly above the ground-state energy $E_0$; then the portion of the configuration space with potential energy $\Phi\le E$ is given by the equation
\begin{equation}
E\ge E_0+\frac{\lambda_1}{2}x_1^2+\frac{\lambda_2}{2}x_2^2+\cdots +\frac{\lambda_{d(N-1)}}{2}x_{d(N-1)}^2,
\label{energy}
\end{equation}
where $x_1$, $x_2$, $\cdots$, $x_{d(N-1)}$ are the deformations along each eigenvectors of the Hessian matrix. Equation (\ref{energy}) specifies a $d(N-1)$-dimensional ellipsoid, for which the hypervolume is
\begin{equation}
\mathcal V_{\mbox{vibrational}}=\frac{\pi^{d(N-1)/2}}{\Gamma[1+d(N-1)/2]} \prod_{j=1}^{d(N-1)} \sqrt{\frac{2\delta E}{\lambda_j} },
\label{vibvolume}
\end{equation}
where $\frac{\pi^{d(N-1)/2}}{\Gamma[1+d(N-1)/2]}$ is the volume of a $d(N-1)$-dimensional hypersphere of unit radius and $\delta E = E-E_0$.

To obtain the total volume of the configuration space for which $\Phi\le E$, one needs to multiply Eq.~(\ref{vibvolume}) with a few additional factors to account for trivial symmetry operations. First, there are $d$ independent translations, each contributing a factor of $\sqrt{N}L$, where $L$ is the side length of the simulation box. The  factor $\sqrt{N}$ comes from the fact that translations correspond to diagonal movements in the configuration space. An additional factor, $f$, that depends on the space dimension and the simulation box shape, also needs to be included to account for rotations and reflections. For $d=2$ with square box, $f=8$, since such boxes allow rotations of $0^\circ$, $90^\circ$, $180^\circ$, and $270^\circ$, and a combination of any rotation with a chirality inversion. Lastly, particle permutations contribute a factor of $N!$. Overall, the total volume in the configuration space is
\begin{equation}
\mathcal V= N^{d/2} V f N! \frac{\pi^{d(N-1)/2}}{\Gamma[1+d(N-1)/2]} \prod_{j=1}^{d(N-1)} \sqrt{\frac{2\delta E}{\lambda_j} },
\end{equation}
where $V=L^d$ is the volume of the simulation box.

The density of states is the surface area of the total volume in the configuration space for which $\Phi\le E$, and is therefore the derivative of $\mathcal V$ to $E$:
\begin{equation}
\begin{split}
g(E)=\frac{d \mathcal V}{d E}= N^{d/2} V f N! \frac{\pi^{d(N-1)/2}}{\Gamma[1+d(N-1)/2]} \\ \times\left(\prod_{j=1}^{d(N-1)} \sqrt{\frac{2}{\lambda_j} }\right) \frac{d(N-1)}{2} \delta E^{d(N-1)/2-1}.
\end{split}
\end{equation}

\subsection{Density of states, $g(E)$, from Wang-Landau Monte Carlo simulations}

We now use the WLMC algorithm to calculate $g(E)$ for the perfect-glass system. To do so, we first divide the energy range $E_0\le \Phi < \ee{5}$ into $N_{\mbox{bin}}=2\e{4}$ bins that are equidistant in a logarithmic scale. Let the minimum and maximum energies of a bin be $E_{\mbox{min}}$ and $E_{\mbox{max}}$, the WLMC algorithm allows one to calculate $\displaystyle g_{\mbox{bin}}= c \int_{E_{\mbox{min}}}^{E_{\mbox{max}}}g(E)dE$ over every bin, where $c$ is an unknown constant independent of the bin \cite{wang2001efficient}. We then determine $c$ by the condition $ V^N=\displaystyle \int_{E_0}^{\infty}g(E)dE = \sum g_{\mbox{bin}}$, where the upper limit of the integration can be replaced with $\ee{5}$, since $g(E)$ turns out to be negligible for very large $E$. We finally divide $\displaystyle g_{\mbox{bin}}$ with $(E_{\mbox{max}}-E_{\mbox{min}})$ to find out $g(E)$ at each bin.

We perform a total of $1500$ stages of Monte Carlo simulations, each consisting of $N_{\mbox{trial}}=4\e{7}$ trial moves.
In each trial move, a random particle is moved by a distance of $xyL$ in every direction, where $x$ is uniformly distributed between -1 and 1 and $y$ has a $50\%$ probability of being 0.2 and a  $50\%$ probability of being 0.002. 
The WLMC algorithm has a tuning parameter, called the ``modification factor'' in \cite{wang2001efficient}, that affects its efficiency and accuracy. Following Ref.~\onlinecite{belardinelli2007wang}, we let this factor be $f=\exp\{\max[ 2N_{\mbox{bin}}/(N_{\mbox{trial}}i), \exp(-0.1i) ]\}$ at the $i$th stage, where $\max(a, b)$ denotes the maximum value between $a$ and $b$.

\begin{figure}
\includegraphics[width=0.49\textwidth]{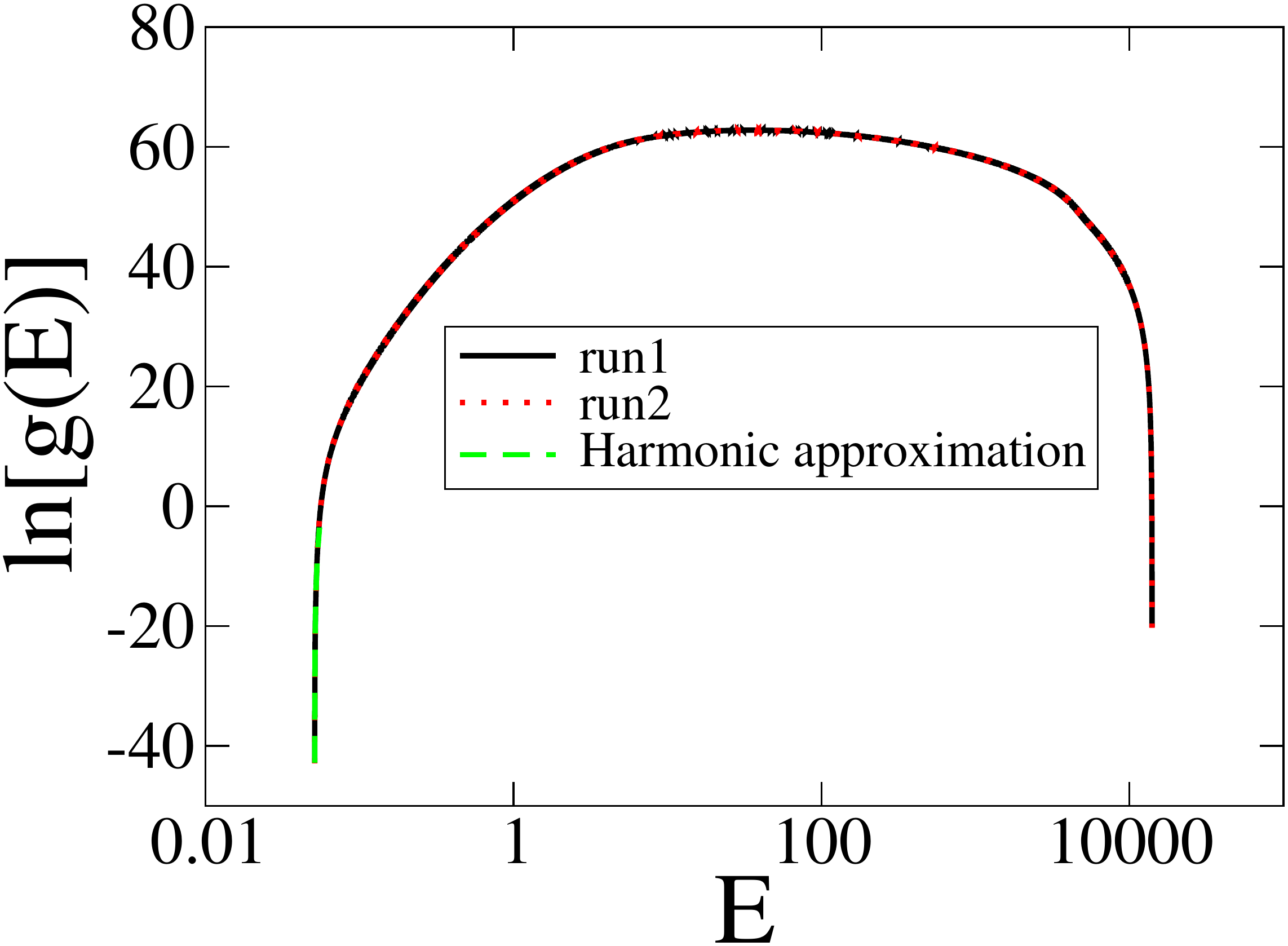}

\vspace{0.3in}

\includegraphics[width=0.49\textwidth]{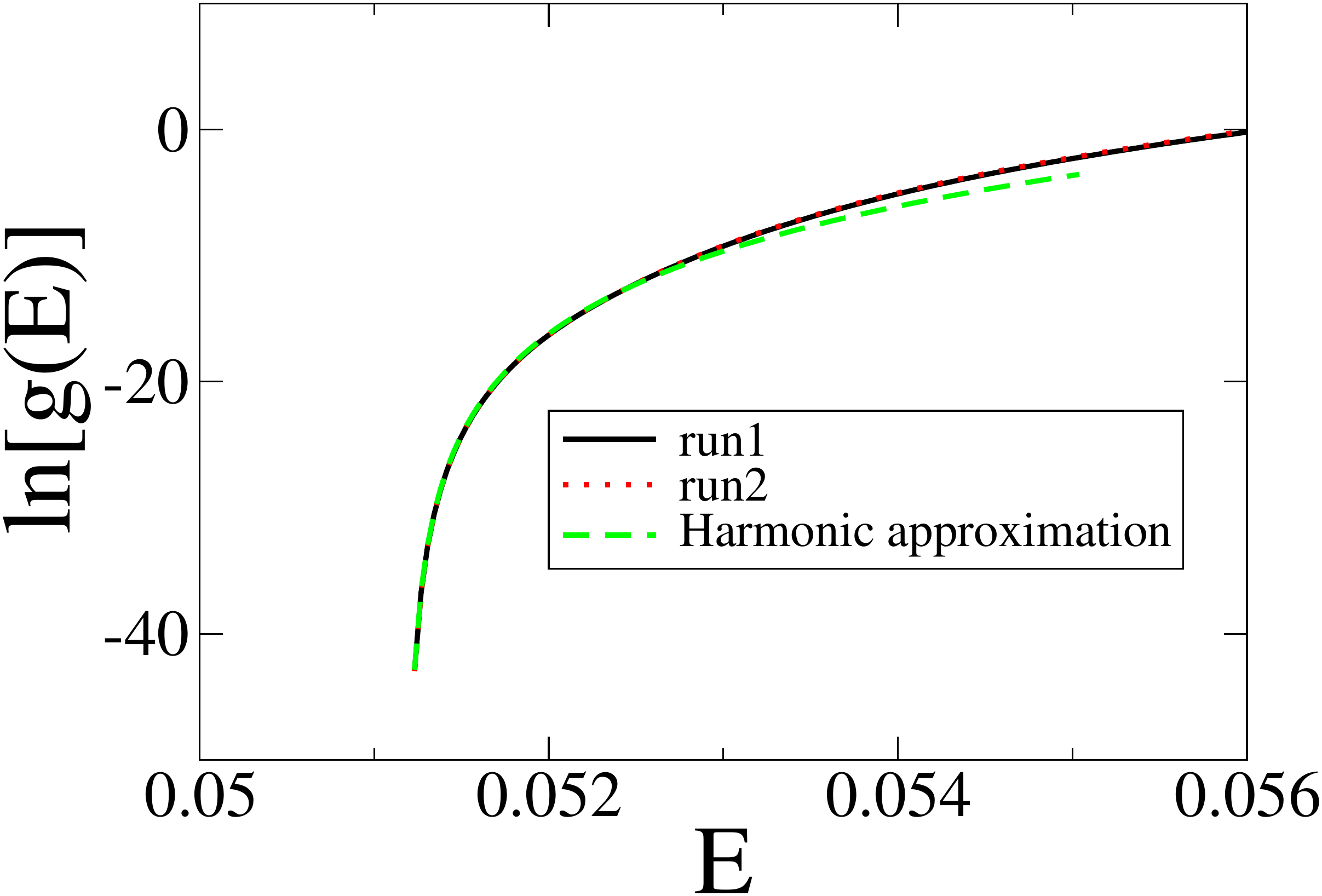}
\caption{(Top) Natural logarithm of the density of states, $g(E)$, from two independent runs of WLMC simulations, and from the harmonic approximation. (Bottom) A zoomed-in view near the ground-state energy $E_0=0.0512129\ldots$.}
\label{fig:WLMC}
\end{figure}

We have performed two independent runs of the simulations detailed above. The resulting $g(E)$ is presented in Fig.~\ref{fig:WLMC} and compared with the $g(E)$ obtained from the harmonic approximation. At the lowest energies, $g(E)$ from both runs agree very well with that from the harmonic approximation, differing by less than $12\%$.
If the ground state was two-fold degenerate, there would be a two-fold difference between the calculated $g(E)$. 
This verifies the uniqueness of the perfect-glass ground state.

\begin{figure}
\includegraphics[width=0.32\textwidth]{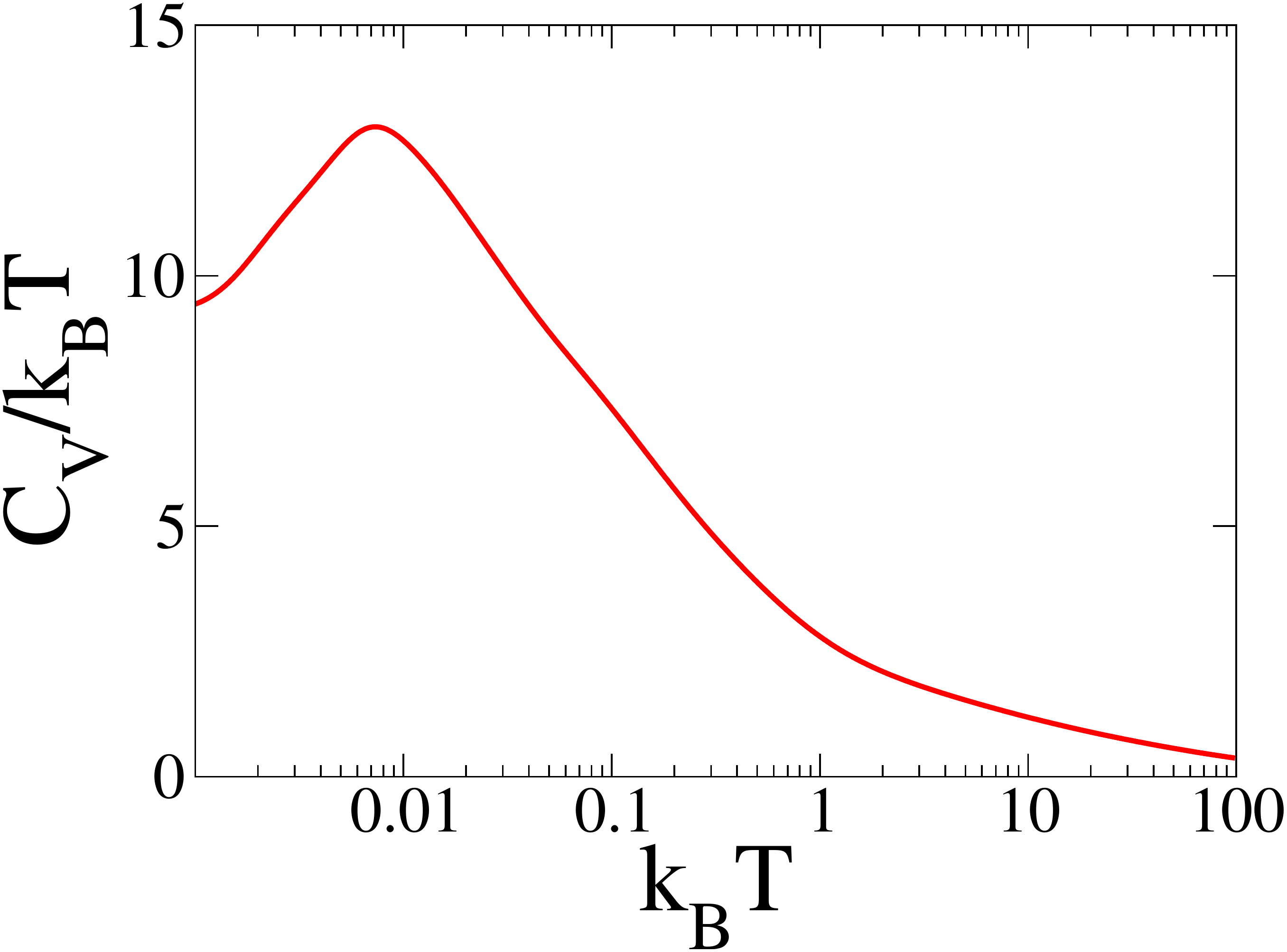}
\caption{Isochoric heat capacity $C_V$ of the perfect-glass system of $d=2$, $\alpha=1$, $\chi=1.89$, and $N=10$. Here, the constant contribution to the heat capacity from the kinetic energy, $C_{V \mbox{, kinetic}}=dN/2$, is excluded.}
\label{fig:Cv}
\end{figure}

From the density of states, we have also calculated the excess isochoric heat capacity $C_V$ of the system, which is given by
\begin{equation}
C_V=\frac{d\langle \Phi\rangle}{dk_BT},
\end{equation}
where
\begin{equation}
\langle\Phi\rangle=\frac{\int E g(E)\exp(-E/k_BT) dE}{\int g(E)\exp(-E/k_BT) dE}.
\end{equation}
The heat capacity, presented in Fig.~\ref{fig:Cv}, starts at the harmonic value at $T=0$, and begins to rise because the shape of the potential energy landscape is such that the effective harmonic force constants are reduced in order to produce transition pathways to neighboring minima. The reduction of the local effective force constants increases the amount of configuration spaces associated with that particular level of the potential energy, and therefore increases the heat capacity. Eventually, $C_V$ levels off and decreases because the energy landscape becomes irrelevant at very high temperature.

\section{Conclusions and discussion}
To summarize, all previously known disordered classical ground states are caused by interactions that allow continuous configurational deformations without energy change. These deformations also cause the ground state to possess large and extensive entropy. 
In contrast to these previous investigations, here we create disordered classical ground states by penalizing crystalline order, causing no ground-state degeneracy. 
These zero-entropy ground states are in sharp contrast with zero-entropy crystalline ground states, since the latter possess very high symmetry and long-range translational and rotational order.

Our discovery of unique disordered ground states impinge on the famous Kauzmann glass paradox \cite{kauzmann1948nature} and the associated ``ideal glass'' \cite{debenedetti2001supercooled}. 
Historically, it was found that at some low but positive temperature, called the ``Kauzmann temperature,'' the extrapolated entropy of a supercooled liquid would be equal to and then apparently decline below that of the crystal, resulting in what has been called the ``Kauzmann paradox.'' This impossible scenario constitutes an entropy crisis. One resolution is to postulate that as the Kauzmann temperature is reached, the supercooled liquid must undergo a phase transition into an ``ideal glass,'' which is a glass with zero configurational entropy.
This scenario is in contradistinction to the perfect-glass model, which does not require that the presence of a crystal and its thermodynamic properties has to provide a constraint on the behavior of the amorphous manifold of the configurations. Indeed, perfect-glass ground states can never be crystalline nor quasicrystalline.
On the other hand, perfect-glass ground states and ``ideal glasses'' share one common feature: both states are disordered while having zero enumeration entropy.

We would like to stress that while the perfect-glasses interaction is not yet achievable in practice, it is an idealization that is nonetheless valuable because it teaches us what types of many-body molecular interactions are required to attain ``unique'' disorder and hence provides guidance to experimentalists to approximate such interactions in the laboratory with, for example, polymers \cite{zhang2016perfect}.

Finally, we also expect our results to be useful in cryptography, where pseudo-random functions with tunable computational complexity are desired; for example, in deriving an encryption key from a password \cite{krawczyk2010cryptographic}. The task of finding a perfect-glass ground state suits this need, since its complexity can easily be tuned by changing a set of parameters ($d$, $N$, $\alpha$, and $\chi$).

\begin{acknowledgments}

We are very grateful to Giorgio Parisi for helpful discussions.
This work was supported in part by the National Science Foundation (NSF) under Award No. DMR-1714722.
\end{acknowledgments}

\appendix
\section{Summary of Results of the  Numeration Studies}

In Table~\ref{my-label}, we summarize major results obtained from our enumeration studies  for the different
 parameter combinations ($d$, $\alpha$, $\chi$, and $N$) that we  used.

\renewcommand{\arraystretch}{0.65}
\begin{longtable*}{|c|c|c|c|c|c|c|c|c|}
\caption{List of all the parameter combinations ($d$, $\alpha$, $\chi$, and $N$) that we employed to carry out
the  enumeration studies and a summary of results for each combination, which includes the number of inherent structures that we generated, the number of times the ground state structure was achieved, the ground-state energy, the mean energy of inherent structures, and the number of distinct energy levels of inherent structures found.}\\
\label{my-label}\\
\hline
$d$ & $\alpha$ & $\chi$  & $N$  & \parbox{3cm}{Number of inherent structures generated} & \parbox{3cm}{Number of lowest-energy structures generated} & \parbox{2cm}{Lowest energy} & \parbox{1.9cm}{Mean energy (of inherent structures)} & \parbox{2cm}{Number of distinct energies found} \\ \hline
2 & 1     & 2.20  & 6  &$\ee{6}$                                 & 97510                                        & 0.0534        & 0.094867                             & 37                                \\
2 & 0.5   & 1.89 & 10 &$3\e{7}$                               & 1769                                         & 0.0059314     & 0.046363                             & 34719                             \\
2 & 1     & 1.89 & 10 &$3\e{7}$                                                                       & 14442                                        & 0.0512129     & 0.126145                             & 7398                              \\
2 & 2     & 1.89 & 10 &$3\e{7}$                                                                       & 1508436                                      & 0.835746      & 0.953847                             & 339                               \\
2 & 6     & 1.89 & 10 &$3\e{7}$                                                                       & 5178002                                      & 2.73031       & 2.86945                              & 34                                \\
2 & 1     & 1.87 & 16 &$\ee{8}$                                         & 147                                          & 0.0618558     & 0.178029                             &      $2.16\e{7}$                             \\
2 & 1     & 1.89 & 20 &$\ee{9}$                                                                                 & 40                                           & 0.0664875     & 0.204629                             &$7.41\e{8}$                                                               \\
2 & 6     & 1.89 & 20 &$\ee{8}$                                                                                 & 265084                                       & 5.37199       & 5.74988                              & 147590                            \\
2 & 6     & 1.90  & 30 &$\ee{8}$                                                                                 & 1634                                         & 8.0647        & 8.57543                              &$4.08\e{7}$                                                               \\
2 & 6     & 1.87 & 40 &$\ee{8}$                                                                                 & 11                                           & 10.6843       & 11.3862                              &$9.10\e{7}$                                                                                                 \\
1 & 1     & 1.79 & 20 &$2\e{7}$                                         & 60490                                        & 0.517475      & 1.015135                             & 3492                              \\
1 & 1     & 1.74 & 40 &$5\e{7}$                                         & 24                                           & 0.991197      & 1.849102                             &$3.06\e{7}$                                                                                                 \\
1 & 1     & 1.79 & 50 &$\ee{9}$                                                                                                                         & 12                                           & 1.68337       & 2.94643                              &$8.45\e{8}$                                                                                                 \\
1 & 6     & 1.75 & 60 &$\ee{8}$                                         & 28                                           & 77.8601       & 79.7294                              &$5.46\e{7}$                                                                                                 \\
1 & 6     & 1.75 & 70 &$\ee{9}$                                         & 27                                           & 93.3095       & 95.7188                              &$5.94\e{8}$                                                                                                 \\
1 & 1     & 2.00    & 10 &$\ee{7}$                                         & 739724                                       & 0.866727      & 1.31398                              & 20                                \\
1 & 1     & 2.00    & 11 &$\ee{7}$                                                                                 & 651397                                       & 0.929444      & 1.42011                              & 24                                \\
1 & 1     & 2.00    & 12 &$\ee{7}$                                                                                 & 589273                                       & 1.2673        & 1.69868                              & 48                                \\
1 & 1     & 2.00    & 13 &$\ee{7}$                                                                                 & 242182                                       & 1.31007       & 1.87405                              & 78                                \\
1 & 1     & 2.00    & 14 &$\ee{7}$                                                                                 & 358037                                       & 1.53538       & 2.06303                              & 109                               \\
1 & 1     & 2.00    & 15 &$\ee{7}$                                                                                 & 142763                                       & 1.61837       & 2.29402                              & 201                               \\
1 & 1     & 2.00    & 16 &$\ee{7}$                                                                                 & 158544                                       & 1.69043       & 2.50381                              & 351                               \\
1 & 1     & 2.00    & 17 &$\ee{7}$                                                                                 & 105541                                       & 1.85802       & 2.74284                              & 557                               \\
1 & 1     & 2.00    & 18 &$\ee{7}$                                                                                 & 65853                                        & 1.98903       & 2.96204                              & 959                               \\
1 & 1     & 2.00    & 19 &$\ee{7}$                                                                                 & 27438                                        & 2.31345       & 3.17478                              & 1578                              \\
1 & 1     & 2.00    & 20 &$\ee{7}$                                                                                 & 22617                                        & 2.40054       & 3.4008                               & 2613                              \\
1 & 1     & 2.00    & 21 &$\ee{7}$                                                                                 & 15771                                        & 2.55858       & 3.64033                              & 4527                              \\
1 & 1     & 2.00    & 22 &$\ee{7}$                                                                                 & 20318                                        & 2.63018       & 3.85341                              & 7645                              \\
1 & 1     & 2.00    & 23 &$\ee{7}$                                                                                 & 9110                                         & 2.80526       & 4.06194                              & 12665                             \\
1 & 1     & 2.00    & 24 &$\ee{7}$                                                                                 & 5778                                         & 2.93012       & 4.289                                & 21383                             \\
1 & 1     & 2.00    & 25 &$\ee{7}$                                                                                 & 1754                                         & 3.1362        & 4.52102                              & 36116                             \\
1 & 1     & 2.00    & 26 &$\ee{7}$                                                                                 & 2095                                         & 3.21506       & 4.7385                               & 60728                             \\
1 & 1     & 2.00    & 27 &$\ee{7}$                                                                                 & 1261                                         & 3.45142       & 4.95513                              & 100960                            \\
1 & 1     & 2.00    & 28 &$\ee{7}$                                                                                 & 1618                                         & 3.60496       & 5.1805                               & 168599                            \\
1 & 1     & 2.00    & 29 &$\ee{7}$                                                                                 & 573                                          & 3.84163       & 5.39782                              & 275767                            \\
1 & 1     & 2.00    & 30 &$\ee{7}$                                                                                 & 518                                          & 4.01308       & 5.60469                              & 442279                            \\
1 & 6     & 2.00    & 10 &$\ee{7}$                                                                                 & 2715392                                      & 14.13106      & 14.2772                              & 5                                 \\
1 & 6     & 2.00    & 11 &$\ee{7}$                                                                                 & 2749262                                      & 16.47931      & 16.8239                              & 13                                \\
1 & 6     & 2.00    & 12 &$\ee{7}$                                                                                 & 1432882                                      & 19.08413      & 19.325                               & 15                                \\
1 & 6     & 2.00    & 13 &$\ee{7}$                                                                                 & 470326                                       & 21.22258      & 21.5804                              & 13                                \\
1 & 6     & 2.00    & 14 &$\ee{7}$                                                                                 & 624508                                       & 23.95933      & 24.2088                              & 22                                \\
1 & 6     & 2.00    & 15 &$\ee{7}$                                                                                 & 578671                                       & 26.13922      & 26.7836                              & 43                                \\
1 & 6     & 2.00    & 16 &$\ee{7}$                                                                                 & 554290                                       & 28.93276      & 29.444                               & 59                                \\
1 & 6     & 2.00    & 17 &$\ee{7}$                                                                                 & 196175                                       & 31.03469      & 31.902                               & 75                                \\
1 & 6     & 2.00    & 18 &$\ee{7}$                                                                                 & 471502                                       & 34.10818      & 34.6142                              & 102                               \\
1 & 6     & 2.00    & 19 &$\ee{7}$                                                                                 & 224480                                       & 36.54381      & 37.2612                              & 154                               \\
1 & 6     & 2.00   & 20 &$\ee{7}$                                                                                 & 214371                                       & 39.26553      & 39.9763                              & 204                               \\
1 & 6     & 2.00    & 21 &$\ee{7}$                                                                                 & 73660                                        & 41.83513      & 42.6645                              & 301                               \\
1 & 6     & 2.00    & 22 &$\ee{7}$                                                                                 & 98901                                        & 44.40762      & 45.3614                              & 422                               \\
1 & 6     & 2.00    & 23 &$\ee{7}$                                                                                 & 69845                                        & 47.1069       & 48.1436                              & 677                               \\
1 & 6     & 2.00    & 24 &$\ee{7}$                                                                                 & 57822                                        & 49.86945      & 50.8796                              & 887                               \\
1 & 6     & 2.00    & 25 &$\ee{7}$                                                                                 & 57726                                        & 52.37915      & 53.5622                              & 1343                              \\
1 & 6     & 2.00    & 26 &$\ee{7}$                                                                                 & 26194                                        & 55.20688      & 56.4128                              & 1966                              \\
1 & 6     & 2.00    & 27 &$\ee{7}$                                                                                 & 22438                                        & 57.52615      & 59.1253                              & 2853                              \\
1 & 6     & 2.00    & 28 &$\ee{7}$                                                                                 & 12874                                        & 60.70169      & 61.9653                              & 4022                              \\
1 & 6     & 2.00    & 29 &$\ee{7}$                                                                                 & 2699                                         & 62.57372      & 64.6753                              & 5898                              \\
1 & 6     & 2.00    & 30 &$\ee{7}$                                                                                 & 10039                                        & 66.15001      & 67.5981                              & 8796                              \\
3 & 1     & 1.70  & 10 &$3\e{7}$                                                                                                               & 1418                                         & 0.0020304     & 0.0196315                            &$1.04\e{6}$                                                                                                 \\
3 & 6     & 1.77 & 20 &$3\e{7}$                                                                                                               & 553282                                       & 1.05579       & 1.24064                              & 945314                            \\
3 & 6     & 1.75 & 30 &$3\e{7}$                                                                                                               & 518                                          & 1.69167       & 1.88439                              &$2.51\e{7}$                                                                                                
\\ \hline
\end{longtable*}

\section{Details of the Configuration Comparison Algorithm}
Here provide details of the algorithm that we devised to compare two configurations to determine whether 
one of them can be superposed onto to the other after a translation, a rotation, and/or a reflection. 

We begin with a description for the one-dimensional case for simplicity. For each configuration, we find a ``characteristic vector'' by the following steps:
\begin{itemize}
\item Find the closest pair of particles, $A$ and $B$. Find out their locations, $\mathbf r_A$ and $\mathbf r_B$.
\item Find the distance from particle $A$ to its second closest neighbor particle, $d_A$; and the same distance for particle $B$, $d_B$.
\item If $d_A>d_B$, then swap particles $A$ and $B$.
\item The characteristic vector is $\mathbf v_1=\mathbf r_B-\mathbf r_A$. 
\end{itemize}
The characteristic vector is invariant to configuration translations and particle permutations, and rotates or reflects if the configuration is rotated or reflected.
Thus, if the two configurations are indeed related to each other through these trivial symmetry operations, then their characteristic vector must be related to each other by a constant 1 or -1, {\it i.e.,}
\begin{equation}
\mathbf v_1^2= R \mathbf v_1^1,
\end{equation}
where $\mathbf v_1^j$ is the characteristic vector of the $j$th configuration, and $R$ is either 1 or -1. If $R=1$, then the two configurations are not related to each other by any rotation or reflection. If $R=-1$, then the two configurations are related to each other by a $180^\circ$ rotation, or equivalently in one dimension, a reflection.
The translation relating the two configurations can be found by the difference of the location of particle $A$: $\mathbf t = \mathbf r_A^2-R \mathbf r_A^1$, where the superscripts indicate different configurations.
Having found the translation and rotation relating these configurations, one can verify that for each particle $j$ in the first configuration, at location $R \mathbf r_j^1+\mathbf t$ there is a particle in the second configuration. If so, and if the two configurations have the same number of particles, then these two configurations must be related to each other through symmetry operations.

To generalize this method to $d>1$ dimensions, one must find $d$ characteristic vectors, derived from $d$ closest particle pairs. Solving the following matrix equation gives the rotation and/or reflection matrix between the two configurations, $R$.
\begin{equation}
\begin{pmatrix}& & & \\ \mathbf v_1^2&\mathbf v_2^2&\cdots&\mathbf v_d^2\\ & & & 
\end{pmatrix}=R
\begin{pmatrix}& & & \\ \mathbf v_1^1&\mathbf v_2^1&\cdots&\mathbf v_d^1\\ & & & 
\end{pmatrix}
\label{p41}
\end{equation}
where $\mathbf v_i^j$ is the $i$th characteristic vector of the $j$th configuration. The translation relating the two configurations can be found  similarly by $\mathbf t = \mathbf r_A^2-R \mathbf r_A^1$, where $\mathbf r_A^j$ denotes the starting particle in finding the first characteristic vector in configuration $j$. Similar to the 1D case, the $j$th particle in configuration 1 still corresponds to a particle at $R \mathbf r_j^1+\mathbf t$ in configuration 2.

\end{document}